\documentclass[%
 aip,
 jmp,%
 amsmath,amssymb,
 reprint,%
]{revtex4-1}

\usepackage{graphicx}
\usepackage{dcolumn}
\usepackage{bm}
\begin{document}

\preprint{AIP/123-QED}
\title{Interplay of Delay and multiplexing: Impact on Cluster Synchronization}

\author{Aradhana Singh}%
\affiliation{Complex Systems Lab, Discipline of Physics, Indian Institute of Technology Indore,
Khandwa Road, Simrol, Indore-453552}
\author{Sarika Jalan}%
 \email{sarikajalan9@gmail.com; Corresponding Author}
\affiliation{Complex Systems Lab, Discipline of Physics, Indian Institute of Technology Indore,
Khandwa Road, Simrol, Indore-453552}
\affiliation{Centre for Bio-Science and Bio-Medical Engineering, Indian Institute of Technology Indore, Khandwa Road, Simrol, Indore-453552}
\author{Stefano Boccaletti}%
\affiliation{CNR-Institute of Complex Systems, Via Madonna del Piano, 10, 50019 Sesto Fiorentino, Florence, Italy}
\affiliation{The Embassy of Italy in Tel Aviv, 25 Hamered street, 68125 Tel Aviv, Israel}

\date{\today}

\begin{abstract}
Communication delays and multiplexing are ubiquitous features of real-world networked systems. We here introduce a simple model where these two
features are simultaneously present, and report the rich phenomenology which is actually due to their interplay on cluster synchronization.
A delay in one layer has non trivial impacts on the collective dynamics of the other layers, enhancing or suppressing synchronization.
At the same time, multiplexing may also enhance cluster synchronization of delayed layers.
We elucidate several nontrivial (and anti-intuitive) scenarios, which are of interest 
and potential application in various real-world systems, where the introduction of a delay may render synchronization of a layer robust against changes in the properties of the other layers.  
\end{abstract}
\maketitle

\begin{quotation}
Synchronization plays a vital role in proper functioning of the brain, social, ecological and many
other complex real-world systems. These real-world systems possess multiple relations
among its units (nodes) and thus can be represented by a multiplex network.
Furthermore, communication delays naturally
arise in real-world systems, and they may affect either a single layer of all the layers of the network.  We introduce a simple model where two features are simultaneously present,
and elucidate several phenomena of cluster synchronization due
to their interplay. Interestingly,
cluster synchronization of a layer in a multiplex network can be tuned by the delay
value in the other layers. Furthermore, by introducing delay,
cluster synchronization in one layer can be made robust
against changes in the delay value as well as network architecture of the other layer.
These results are important for multiplex systems with more than one type of vital
resources, as for instance water supply and power plants of a city,
where controlling synchronizability is a task of fundamental importance.
\end{quotation}

Synchronization is one of the most common process through which real-world systems (such as chirping crickets, or flashing fireflies \cite{book_Kurths}) act synergically, and it is a crucial feature which facilitates the collective functioning of man-made systems, like power grids or communication systems \cite{Powergrid}. In particular, cluster synchronization (CS) refers to the case where only few components of the system form clusters of synchronized units \cite{eco_clus}.

\begin{figure}
\centering
\includegraphics[width=0.9\columnwidth,height=4.0cm]{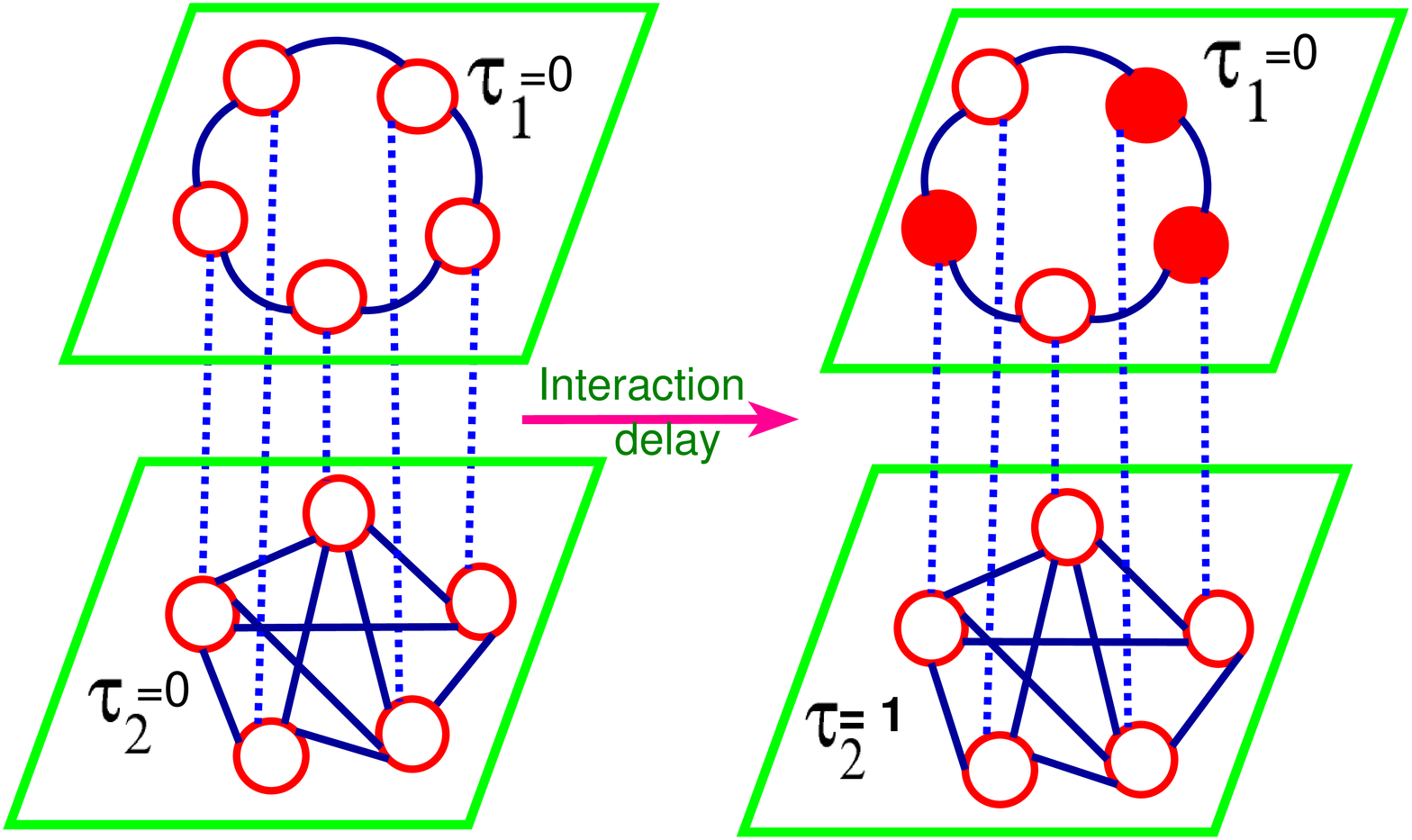}
\caption{(Color online) Sketch of multiplex networks with interaction delays. Solid and dotted lines represent intra and inter-layer connections, respectively. Filled circles denote nodes that are synchronized and form a cluster.}
\label{Fig-multiplex}
\end{figure}

On the other side, multiple interactions co-exist in general among the same set of constituents of a system, which led to a spurt of modeling activities under the so-called  multiplex network framework, where nodes are copied in different layers featuring different shared relationships. An example is a transport network, where cities can form different layers depending on whether they are connected by rail, bus or
air \cite{Multiplex_Bocaletti, lab_multiplex}. Furthermore, layers are usually strongly inter-dependent (for instance, a strike of the bus service may result in overloading the rail and air traffic routes), and major consequences of such dependencies are known to be, for instance, explosive synchronization, diffusion dynamics, emergence of chimera, epidemic spreading, and intra- 
and inter-layer synchronization \cite{Multiplex_Bocaletti, Sapta_chimera_multiplex}. CS, in particular, of sparse networks, gets enhanced upon multiplexing with other sparse networks, whereas it is suppressed upon multiplexing with dense networks \cite{Multiplex_cluster}.

In addition, communication delays naturally arise in real world systems \cite{book_delay}, and they may affect either a single layer or all the layers of the network. Global \cite{Delay_globsyn_multiplex} and breathing synchronization \cite{breathing_syn_multiplex} are phenomena emerging due to delays in the inter-layer coupling, but still, the impact of delay on CS  is not fully investigated. In this Letter, we focus on the simultaneous presence of delay and multiplexing, and examine the impact of their interplay on CS. We will show that delayed interactions in a layer may lead to an enhancement or suppression in CS of other layers.  Our findings will be exemplified for multiplex networks having different combinations of 1-d lattices, scale-free (SF) and Erd\"os-Re\'nyi (ER) random graphs. The used 1-d lattices have circular boundary conditions with each node having exactly $k$ nearest neighbors.
The scale-free networks are constructed using the Barab\'asi-Albert algorithm by starting with
$\langle k \rangle$ nodes and  adding one node with $\langle k \rangle$ connections
at each step. The Erd\"os-R\'enyi networks are constructed by connecting all pairs of
the nodes with a probability $p$. \cite{rev-network}.

\begin{figure}
\centerline{\includegraphics[width=0.8\columnwidth]{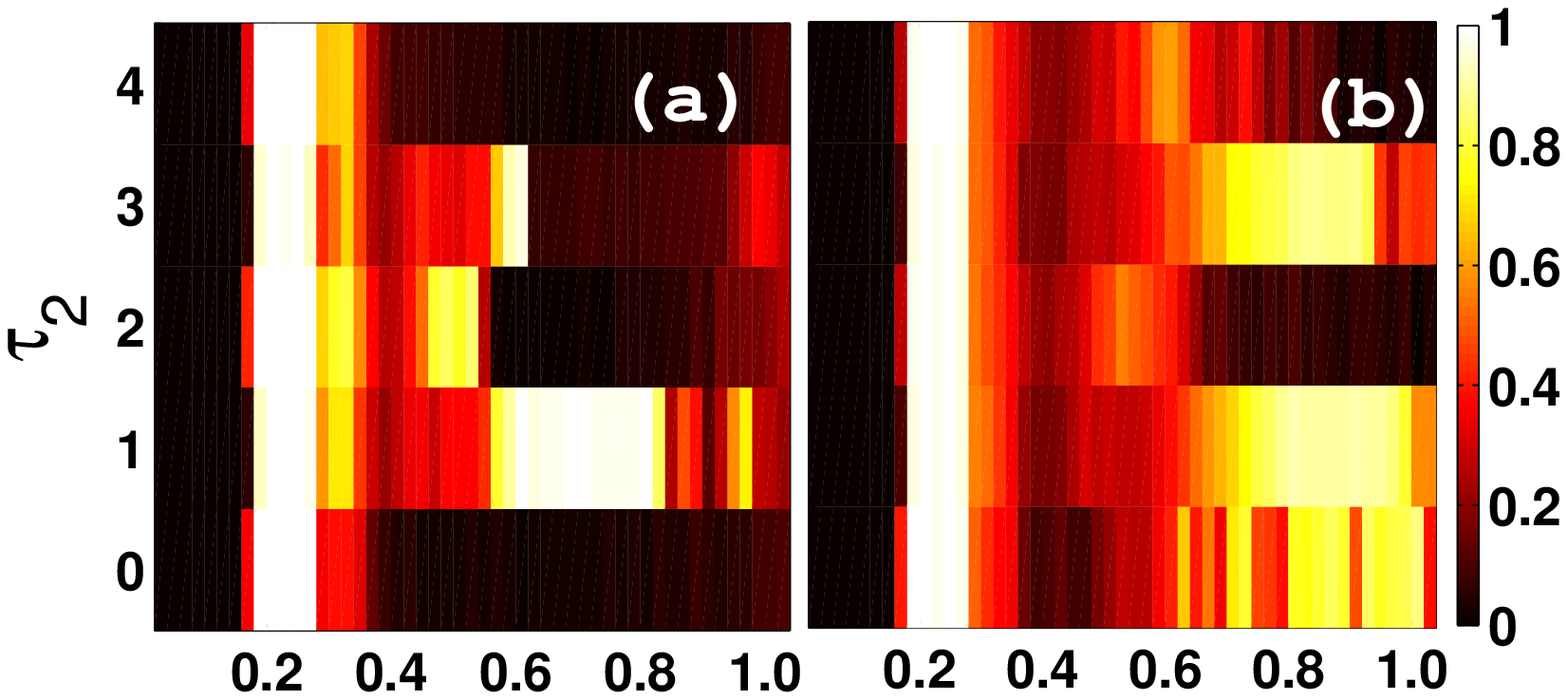}}
\centerline{\includegraphics[width=0.8\columnwidth]{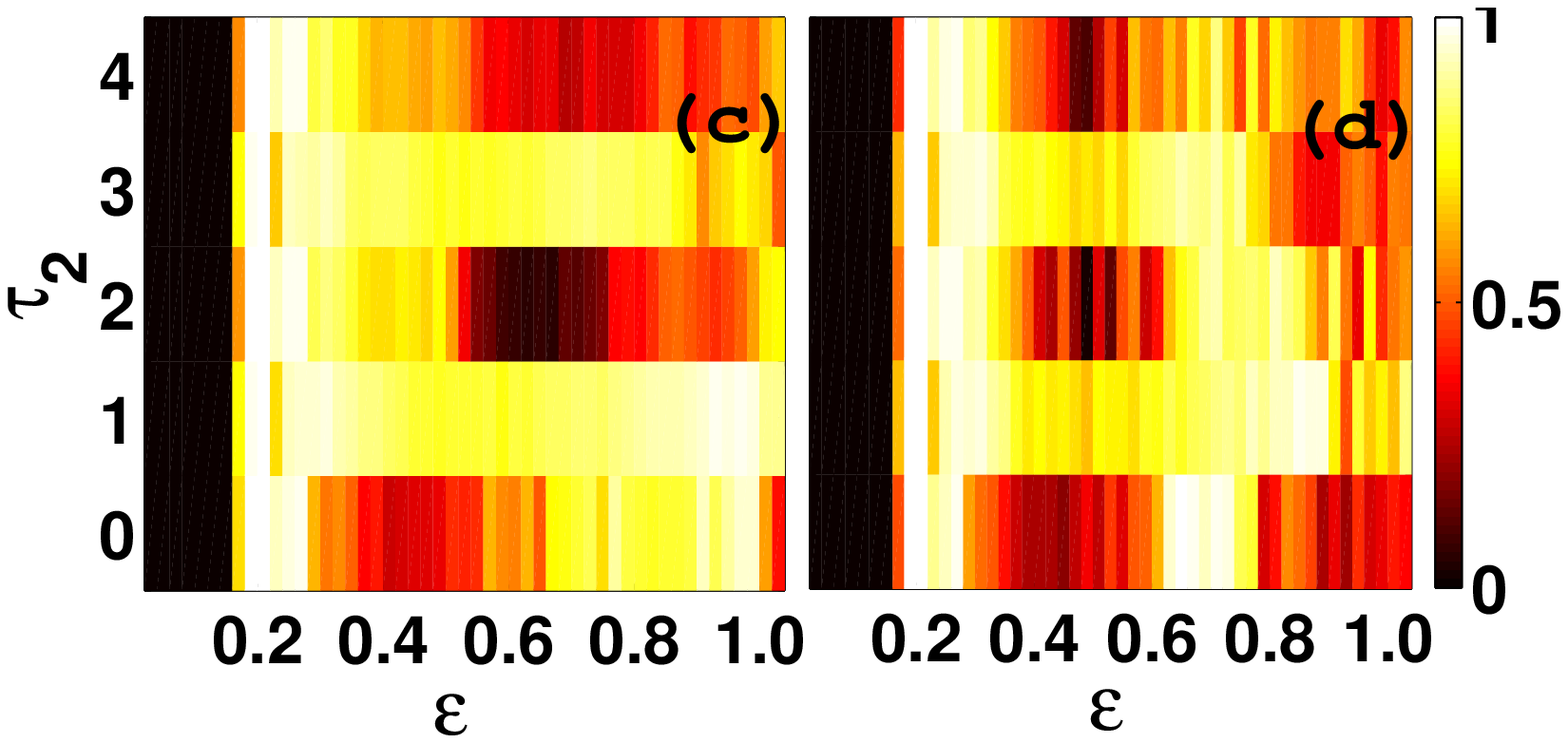}}
\caption{(Color online). $f_{clus}^1$ (see text for definition) in the parameter space ($\tau_2$, $\epsilon$) for (a) a biplex made of two 1-d lattices, (b) a 1-d lattice multiplexed with an ER network, (c) an ER network multiplexed with 1-d lattice, and (d) a SF network multiplexed with 1-d lattice. $N=100$,  $\tau_1=0$, and $\langle k_1 \rangle =\langle k_2 \rangle =4$. Data refer to ensemble averages over 20 different realizations of random initial conditions.
}
\label{Fig_phase}
\end{figure}

Our starting point is a multiplex network (with $N$ nodes in each layer, $l$ layers, and with a number of connections $N_c$). The system is described by a time
dependent vector state ${\bf X}_{i} \equiv \{x_i (t), i=1,2,\hdots,l*N \}$, which evolves following a model of coupled maps \cite{book_delay}:
\begin{eqnarray}
x_i(t+1) = (1-\varepsilon) f(x_i(t)) + \frac{\varepsilon}{k_i} \sum_{j=1}^{Nl} A_{ij} g(x_j(t-\tau_{ij})), \;
\label{cml}
\end{eqnarray}

where $A$ is a supra-adjacency matrix: if only two layers are present in the network, $A= \left( \frac{A^1, I}{I, A^2 } \right)$, where $A^1$ ($A^2$) is the adjacency matrix of layer
$1$ (2). In Eq. (\ref{cml}),  $\varepsilon$ is the overall coupling strength, and $k_i= \sum_{j=1}^{Nl}A_{ij}$ is the degree of the $i^{th}$ node.
The functions $f$ and $g$ are both taken to be the logistic map, so that
$f(x) = g(x) = \mu x(1-x)$, with $\mu=4$.
Coupled logistic map, due to its simplicity yet
ability to display complex dynamical behavior, have been widely investigated to 
understand various complex phenomena manifested by a diverse range of 
real-world systems \cite{logistic_map}.
The delay $\tau_{ij}$ is the communication time that information takes to travel from unit $i$ to unit $j$. We here consider a symmetric ($\tau_{ij}$ = $\tau_{ji}$) and homogeneous ($\tau^1_{ij} = \tau_1$ and $\tau^2_{ij} = \tau_2 \; $ $\forall \; i, j$) case, with
$\tau$ denoting the delay matrix given by $\tau= \left( \frac{\tau^1, 0}{0, \tau^2} \right)$ (see Fig.\ref{Fig-multiplex} for a sketch of the system under study).
Notice that we do not consider the delay in the inter-layer couplings, and only intra-layer communication delay is accounted for.

We focus on phase synchronization, which is here defined as follows \cite{phase_syn_prl998}:
let $n_i$ and $n_j$ denote the number of times at which the
variables $x_i(t)$ and $x_j(t)$ ($t=1, 2, \hdots T$) exhibit local minima (or maxima)
during a long time interval $T$, and let $n_{ij}$ denote
the number of times these local minima (maxima) match with
each other (i.e. occur at the same time). Then, a phase distance between two nodes $i$ and $j$ can be defined as
$d_{ij} = 1 - 2n_{ij}/(n_i + n_j)$.
Two nodes $i$ and $j$ are phase synchronized if $d_{ij}=0$, and all phase synchronized nodes form a cluster of size $N_{clus}$.
Note that the definition of $d_{ij}$ does not involve the network architecture and is 
measured between all the pairs of the nodes irrespective of if there exists a 
connection between a pair of nodes or not.
However, phase distance between a pair of the nodes may change depending upon the 
network architecture, coupling value and the delay value.
CS is said to be enhanced (reduced) if the the fraction of nodes participating in the
cluster formation, $f_{clus} = N_{clus}/N$, increases (decreases). We define robustness in terms of the phenomenon of enhancement or suppression in CS of a layer upon multiplexing with another layer against change in the network parameter of the layer it is multiplexed with.
When the case under consideration is a biplex (a network formed by only two layers), delays can be limited to one of the two layers, or they can affect the interactions in both layers. The first step is then focusing on the case in which only one layer is made of delayed interactions.
Still, two relevant issues arise: (i) how the delay in a layer affects the dynamics of that same layer, and (ii) what is the impact of a delay acting in one layer on the dynamics of the other (un-delayed) layer.

\begin{figure}[t]
\centerline{\includegraphics[width=0.80\columnwidth]{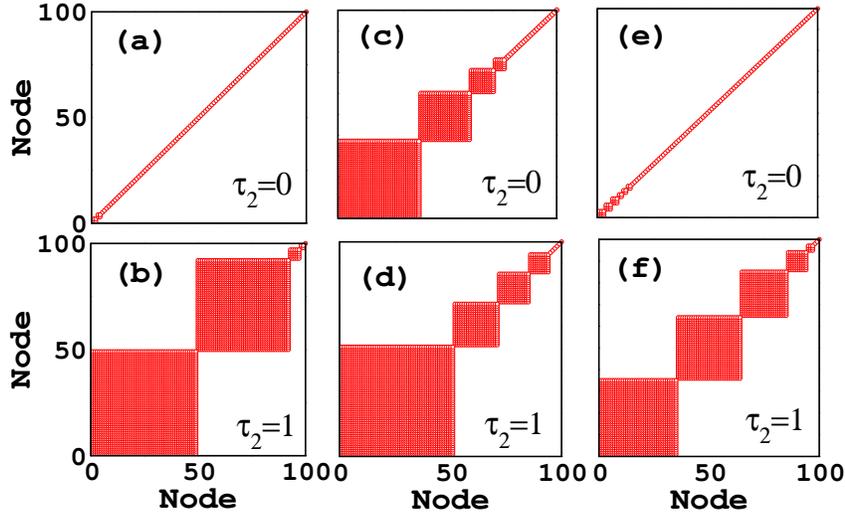}}
\caption{(Color online) Node {\it vs.} node diagram exhibiting enhancement in
CS of the first layer as a delay is introduced in the second layer. 
Circles ($\circ$) represent nodes.
Node numbers are reorganized such that those belonging to the same cluster come consecutively and form a square block. Thus square blocks indicate the phase synchronized clusters, while circles along the diagonal indicate the nodes which are not synchronized with any other node in the network. 
For all the graphs second layer of the multiplex network is represented by the 1-d lattice, whereas the first layer is represented by (a) $\&$ (b) 1-d lattice, (c) $\&$ (d) ER network, (e) $\&$ (f) SF networks. 
Other network parameters are $N_1 = N_2 = 100$, $\langle k_1 \rangle = \langle k_2 \rangle = 4$ and $\varepsilon=0.75$. 
}
\label{Fig_node_node}
\end{figure}

\begin{figure}[t]
\centerline{\includegraphics[width=0.95\columnwidth]{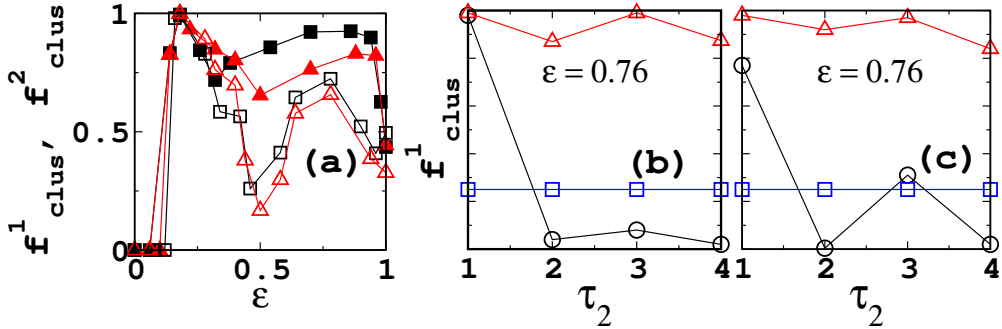}}
\caption{(Color online). (a) $f_{clus}^1$ ($\Box$) and $f_{clus}^2$ ($\triangle$) {\it vs.} $\varepsilon$ for a multiplex 
network consisting of two ER networks with $\tau_1=0$. Closed and open symbols correspond to $\tau_2=1$ and $\tau_2=4$, respectively. 
(b) and (c) $f_{clus}^1$ {\it vs.} $\tau_2$ for multiplex networks comprising (b) 1-d lattice in both  the layers and (c) 1-d lattice in $1^{st}$ layer and SF network in the $2^{nd}$ layer. Squares ($\Box$) correspond to isolated 1-d lattices, whereas 
circles ($\circ$) and triangles ($\triangle$) correspond to un-delayed ($\tau_1=0$) and delayed ($\tau_1=1$) 1-d lattice,  respectively after its multiplexing with a delayed ($\tau_2=1$) network.  Superscript in $f_{clus}$ denoting the layer it is representing.
Same stipulations as in the Caption of Fig. \ref{Fig_phase}.}
\label{Fig-flus_tau}
\end{figure}

We evolve Eq.~\ref{cml} starting from a set of random initial 
conditions and study the phase synchronized clusters after an initial transient. We find that
delayed interactions in a layer may significantly affect the CS of the other layer. The most intriguing case is possibly that of a biplex with regular architecture in both layers, which exhibit very poor CS \cite{Multiplex_cluster}.
For instance,  a 1-d lattice with $\langle k_1 \rangle = 4$ exhibits no CS (at both intermediate and strong couplings) when
multiplexed with another un-delayed 1-d lattice with the same number of connections (see Figs. \ref{Fig_phase}(a) and \ref{Fig_node_node}(a)), whereas it displays about $98\%$ of the nodes taking part in the cluster formation (see Fig. \ref{Fig_phase}(a)) upon multiplexing with a delayed 1-d lattice (for $\tau_2=1$). A further increase in the delay, however, suppresses such an enhancement, and once again no CS is exhibited for $\tau_2 > 4$ at both intermediate and strong couplings (Fig.\ref{Fig-flus_tau}(b)).

A second interesting case is that of a regular network (again a 1-d lattice) multiplexed with a random one (an ER or an SF).
Multiplexing with un-delayed random networks has been shown, indeed, to
enhance CS of a regular network \cite{Multiplex_cluster}. Here, however, the presence of a delay has an
important effect: the multiplexing may lead to a further enhancement (or to a suppression) of CS, depending upon the parity of the delay value.
Namely, odd (even) values of the delay enhance (suppress) CS.
For instance, CS is enhanced in a 1-d lattice with $\langle k_1 \rangle = 4$ upon multiplexing with a un-delayed ER network
of $\langle k_2 \rangle = 4$, it gets further enhanced for $\tau_2=1, 3$, whereas it is suppressed for $\tau_2 = 2, 4$ (see Fig. \ref{Fig_phase}(b)).
Similarly, ER  and SF networks (when multiplexed with a 1-d lattice) exhibit an enhancement in CS for $\tau_2 = 1, 3$, and suppression for $\tau_2 = 2, 4$ (Figs. \ref{Fig_phase}(c), (d)). Figs. \ref{Fig_node_node} reveal that the observed enhancement is significantly more prominent for the 1-d lattice and SF network cases, as compared with the ER network case.

\begin{figure}[t]
\centerline{\includegraphics[width=0.9\columnwidth]{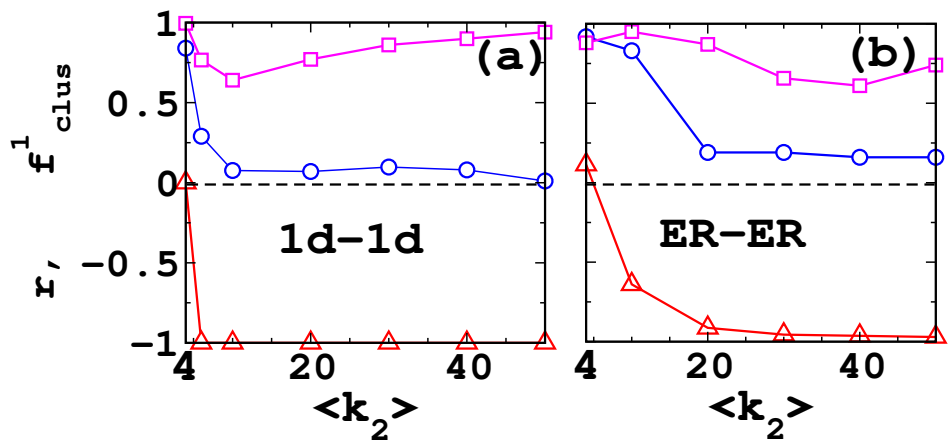}}
\caption{(Color online). Pearson correlation coefficient $r$ (triangles), and  $f_{clus}^1$  (circles for $\tau_1=0$, and squares for $\tau_1=1$)
{\it vs.} $\langle k_2 \rangle$ for (a) a 1-d lattice multiplexed with a 1-d lattice and (b) an ER network multiplexed with another an ER network. $N_1 = N_2 = 100$, $\tau_2 = 1$, $\langle k_1 \rangle = 4$, $\varepsilon = 0.8$. 
}
\label{Fig_random_k2}
\end{figure}

In order to grab the essential reason behind the observed CS enhancement, we notice first that a delay in a layer already enhances CS of that same layer (as also reported for isolated networks \cite{book_delay,Hetero_atay1}). Then, due to multiplexing, such an effect is cooperatively turned into an enhancement of CS in the other, un-delayed, layer. For instance, Fig.~\ref{Fig-flus_tau}(a) shows that the introduction of a delay in an ER network with $\langle k_2 \rangle =4$ conveys more than $60\%$ of the nodes to participate in the cluster formation for  $0.32 \lesssim \varepsilon \lesssim 0.96$ ($0.5 \lesssim \varepsilon \lesssim 0.96$) when $\tau_2=1$ ($\tau_2=4$). In the same Figure, it is seen that the consequence is that the other un-delayed layer (another ER network with $\langle k_1 \rangle =4$), starts exhibiting a good CS in the same coupling regimes.

Such a  cooperation mechanism between the delay and the multiplexing is controlled and regulated by the degree-degree correlations of the replica nodes, calculated using  the Pearson correlation coefficient $r$ \cite{Assor_Newman}. 
The Pearson correlation coefficient $r$ is given by $r = \frac{\sum_{jk} jk(e_{jk} - q_{j} q_{k})}{\sigma_{q}^2}$,
where $e_{jk}$ ($q_k$) is the fraction of nodes of degree $j$ (the fraction of all nodes) connected with nodes of degree $k$, so that $\sum_{jk}e_{jk} = 1$, and $\sum_j e_{jk}= q_{k}$ \cite{Assor_Newman}.

When the delayed layer is denser than the un-delayed one ($\langle k_2 \rangle > \langle k_1 \rangle$), the enhancement in CS of the un-delayed layer is less prominent than the case of both layers having comparable average degrees ($\langle k_1 \rangle \approx \langle k_2 \rangle$). Interestingly, in the  extreme regime (i. e., when $\langle k_2 \rangle > > \langle k_1 \rangle$), CS of the un-delayed layer turns out to be significantly suppressed. As an example, Fig.~\ref{Fig_random_k2}(a) shows that CS of an un-delayed 1-d lattice with $\langle k_1 \rangle = 4$ is maximum upon its multiplexing with another delayed ($\tau_2=1$) 1-d lattice of $\langle k_2 \rangle = 4$, whereas it is completely suppressed when $\langle k_2 \rangle\gtrsim 8$. 
A similar scenario emerges for a pair of ER layers (one delayed, and one un-delayed) representing multiplex network, as depicted by Fig.~\ref{Fig_random_k2}(b). 

The conclusion of this first part of the study is that, in sharp contrast to the case of multiplex networks made of 1-d lattices \cite{Multiplex_cluster}, the delayed evolution of a layer seems to lead (not to lead) to an enhancement in CS of the other un-delayed layer when degree-degree correlations in the replica nodes are neutral (highly disassortative), irrespectively of the topology of each layer.

\begin{figure}[t]
\centerline{\includegraphics[width=1.01\columnwidth]{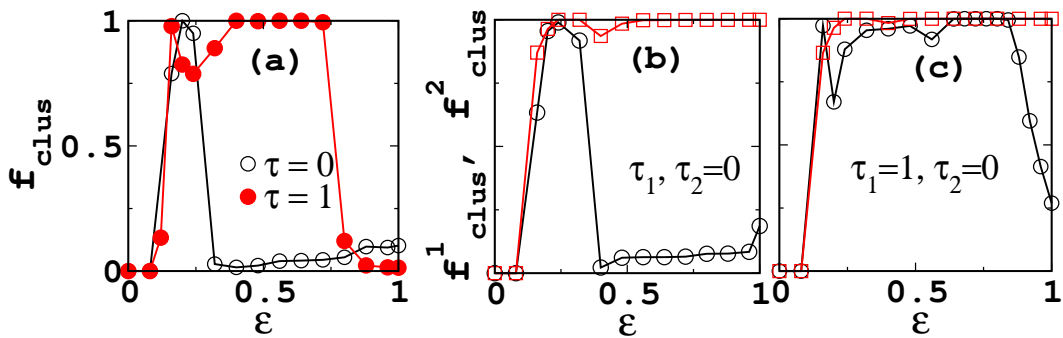}}
\caption{(Color online) $f_{clus}^1$, $f_{clus}^2$ {\it vs.} $\varepsilon$ for (a) a un-delayed (open circles)  and delayed (full circles)
isolated 1-d lattice, (b) a un-delayed 1-d lattice (circles) and a
un-delayed globally connected network (squares) when they are multiplexed with each other, and (c)
a  delayed ($\tau_1=1$) 1-d lattice (circles) multiplexed with an un-delayed globally connected network (squares).
Figures are obtained by averaging over 20 different initial conditions.}
\label{fclus_1d_glob}
\end{figure}

Another relevant scenario is illustrated in Fig. \ref{fclus_1d_glob}. First, in the absence of delay, an isolated 1-d lattice ($\langle k \rangle = 4$) does not exhibit CS for intermediate and strong couplings (see open circles in Fig. \ref{fclus_1d_glob}(a)). Multiplexing it with a un-delayed globally connected network does not lead to enhancement in CS as well, even when $f_{clus}^1\simeq 1$ for the latter network (Fig. \ref{fclus_1d_glob}(b)). When a delay is considered ($\tau=1$) in the initial lattice, CS instead increases: in the range $0.2 \lesssim \varepsilon \lesssim 0.6$ more than $75\%$ of the nodes is participating in the cluster formation,  while CS is almost negligible for $\varepsilon>0.62$ (see the full circles of Fig. \ref{fclus_1d_glob}(a)).
Finally, upon multiplexing, the delayed lattice starts exhibiting an excellent CS in the range $0.14\lesssim \varepsilon \lesssim 0.81$ (with more than $80\%$ of clustered nodes), and even for $\varepsilon>0.82$ $f_{clus}^1$ remains at around 0.3 (see Fig. \ref{fclus_1d_glob}(c)).
This demonstrates a clear enhancement of CS for regular delayed networks, upon multiplexing them with the globally connected networks. Thus, enhancement in CS is the combined impact of multiplexing and delayed coupling.
Similarly, our results indicate that delayed ER networks exhibit strong enhancements in CS upon multiplexing them with regular, ER, and SF networks.

A third scenario is the case in which both layers are affected by communication delays.
The main result is that CS of a delayed layer (say layer 1) remains almost unaffected by changes in the delay value of the second layer.
For instance, a 1-d lattice with coupling delay $\tau_1=1$ exhibits an excellent CS upon multiplexing with another 1-d lattice (or with a SF network) irrespective of the delay value in the other layer [see triangles in Figs.\ref{Fig-flus_tau}(b) and (c)].
In contrast to  the un-delayed case (Fig.~\ref{Fig_phase}), a change in the delay value in one layer does not lead  to drastic changes on CS of another delayed layer.

So far, we focused on the impact of delay in one layer on CS of another (delayed or un-delayed) layer. Next, we show the effect of changing the connection density of a layer on CS of the other delayed layers. Higher connection densities are known to suppress CS in sparse layers \cite{Multiplex_cluster}.
Interestingly, however, the nodes' interactions of a sparse layer are affected by a delay (or if the nodes of both layers have delayed interactions), CS always gets enhanced upon multiplexing, even in the case of maximum disassortativity in the replica nodes (i.e.  for $r=-1$, see curves with squares in Figs. \ref{Fig_random_k2}(a) and (b) for the 1-d lattice and the ER network of $\langle k_1 \rangle=4$).
Notice that the level of CS exhibited by the delayed layer of a multiplex network is much higher than that of the corresponding isolated network. At higher delay values, instead, CS is suppressed.
Therefore,  degree-degree disassortativity in the mirror (replica) nodes does not have effects on CS of the sparse networks if delays affect both layers.
Therefore, in contrast to CS of the un-delayed layer, which is controlled by 
the degree-degree correlations ($r$) of the replica nodes, presence of the 
communication delay makes CS of the delayed layer robust against the change in $r$ 
value.

\begin{figure}[t]
\centerline{\includegraphics[width=1.01\columnwidth]{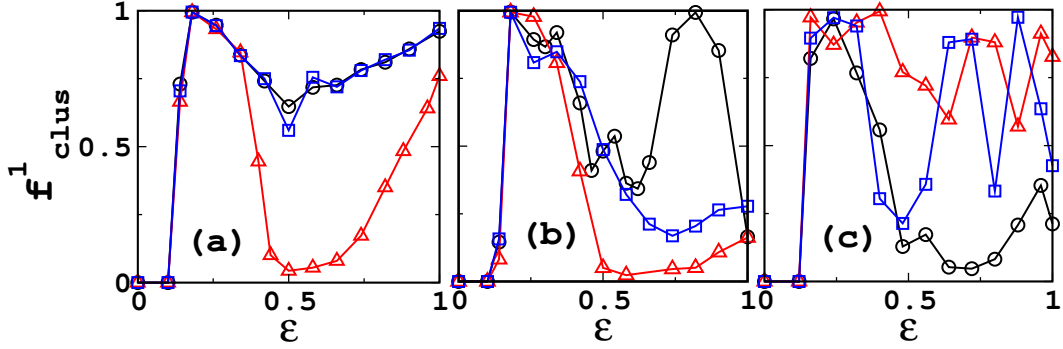}}
\caption{(Color online) $f_{clus}^1$ {\it vs.} $\varepsilon$ in a three-layers multiplex network. (a) an ER layer multiplexed with two other ER layers and (b) a 1-d lattice multiplexed with two other 1-d lattices. In both panels $\langle k_1 \rangle=\langle k_2 \rangle =\langle k_3 \rangle =4$ and  $\tau_3=1$ (circles), 2 (triangles), and  3 (squares).  (c) a 1-d lattice with $\langle k_1 \rangle=4$ multiplexed with other two 1-d lattices with  $\langle k_2 \rangle =\langle k_3 \rangle =20$ for $\tau_1 = \tau_2=0, \tau_3 = 1$ (circles), $\tau_1 =1,  \tau_2 = \tau_3 = 0$ (triangles) and for $\tau_1 =  \tau_3 = 1, \tau_2 =0 $ (squares).  In all layers, $N=100$. Data refer to averages over $100$ random realizations of the initial conditions.
}
\label{Fig-flus_threelayer}
\end{figure}

Finally, in order to demonstrate the generality of our observation, we present results for the case of a three-layers multiplex network, whose supra-adjacency matrix can be written as
\begin{equation}
    A=
      \begin{pmatrix} A^1 & I & I \\ I & A^2 & I \\ I & I & A^3 \end{pmatrix},
\end{equation}
where $A^1$, $A^2$ and $A^3$ are the adjacency matrices corresponding to layers
$1$, $2$ and $3$, respectively. In particular, we  consider the case in which one of the three layers
has delayed interactions (say $\tau_3 \neq 0$) whereas the other two layers are un-delayed ($\tau_1=\tau_2=0$).
The results are reported in Fig.~\ref{Fig-flus_threelayer}. There, a delayed layer consisting of an ER network (a 1-d lattice) is multiplexed with other two un-delayed layers of ER networks (of 1-d lattices). Figs. \ref{Fig-flus_threelayer}(a) and (b) show that CS is strongly enhanced for $\tau_3 = 1, 3$, while it is suppressed for $\tau_3 = 2, 4$. Once again, therefore, the system's behavior is crucially different for odd and even values of the delay time.
A very interesting result is shown in Fig.~\ref{Fig-flus_threelayer}(c): if the delay affects a denser layer, the sparser (and un-delayed) layer do not show enhancements in CS. However, the delayed sparser layer exhibits better CS, irrespective of the delay values in the denser layers. 

\begin{figure}
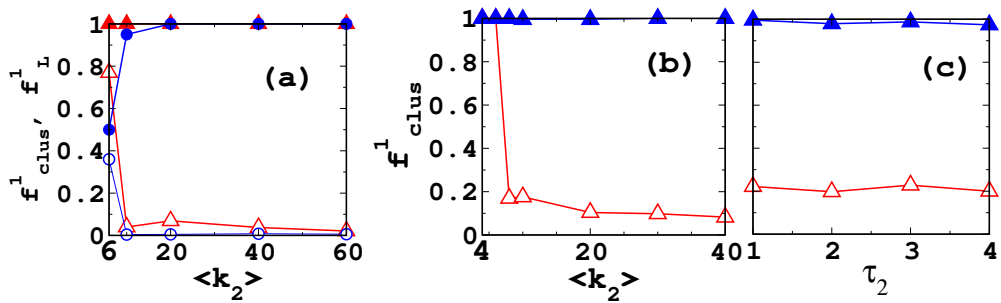

\includegraphics[width=0.34\columnwidth]{Fig8a.eps}
\includegraphics[width=0.6\columnwidth]{Fig8b.eps}
\caption{ (Color online)
Generalization of results presented by Figs.~(\ref{Fig_phase}), (\ref{Fig-flus_tau}) and (\ref{Fig_random_k2}). (a) $f_{clus}^1$ ($\triangle$) and $f_{L}^1$ ($\circ$) {\it vs.} 
$\langle k_2 \rangle$ for ER network (first layer) of $\langle k_1 \rangle = 6$ multiplexed with a 1-d lattice. Here, $N_1 = N_2 =500$. 
Open and closed symbols correspond to un-delayed ($\tau_1=0$) and delayed ($\tau_1=1$) first layer, respectively in a multiplex network with another delayed ($\tau_2=1$) layer. 
(b) $f_{clus}^1$ {\it vs.} $\langle k_2\rangle$ for a multiplex network consisting of two 1-d lattices with non-identical nodes. Other network parameters are $N_1=N_2=100$, $\langle k_1 \rangle = 4$. 
(c) $f_{clus}^1$ {\it vs.} $\tau_2$ for a real-world multiplex network constructed using a celebrity data base \cite{Cam_plos, Cam_epl}. Here, $N_1 = N_2= 241$, $\langle k_1 \rangle = 8$ and $\langle k_2 \rangle = 5$. 
Open and closed triangles ($\triangle$) in (b) and (c) correspond to $f_{clus}^1$ for the un-delayed ($\tau_1=0$) and delayed ($\tau_1=1$) first layer, respectively of a multiplex network having delayed ($\tau_2=1$) second layer.  
 (a) is 
plotted for average over $10$ different realizations of 
the networks and $20$ different random realizations of the initial conditions. (b) and (c) are plotted for average over $20$ random realizations of the initial conditions.
}
\label{Fig_fclus_Fnl_bolly}
\end{figure}

Furthermore, following two phenomena remain unaffected with an increase in the size of the network; (a) CS of an un-delayed layer gets affected with a change in the delay value of the other layer and the degree-degree correlation ($r$) of 
the mirror nodes. (b) In contrast to the un-delayed evolution, CS of a delayed layer is neither governed 
by the degree-degree correlation of the mirror nodes nor is controlled by the 
parameters (such as delay and network architecture) of the other layer. CS of the delayed layer is predominately
governed by its own delay value. For example, Fig.~\ref{Fig_fclus_Fnl_bolly}(a) displays an enhancement in CS of a delayed sparse 1-d lattice of $N_1= N_2 =500$ upon multiplexing. This phenomenon is  irrespective of $r$ of the mirror nodes as 
also observed for smaller size networks ($N_1=N_2=100$) (Fig.~\ref{Fig_random_k2}). Furthermore, the fraction of nodes in the largest cluster ($f_{L}$) exhibits a similar behavior as revealed by $f_{clus}$. 
In real-world systems, not all the nodes 
may be similar, and there may be parameter mismatch in the local dynamical evolution.
To present generality of the phenomena, we present results for the 
coupled non-identical nodes. We consider the coupled non-identical logistic maps \cite {Masoller_nonidenti_map} 
with the parameter $\mu$ uniformly distributed between the range 
$0.38 \lesssim\mu\lesssim 4$. We find that non-identical nodes exhibits 
the similar behavior as observed for the identical nodes. 
Fig.~\ref{Fig_fclus_Fnl_bolly} depicts robustness of the enhancement in CS of 
a delayed layer due to the multiplexing against changes in the connection density of 
the other layer of a biplex network.

Further, we study CS of a real-world 
social network constructed using the celebrity data. The  network is created 
considering actors as the nodes and interaction between a pair of the nodes
is defined based on if they have co-acted in a movie \cite{Cam_plos}.  
The multiplex network is constructed by dividing the nodes into different layers 
depending on their genre \cite{Cam_epl}. A pair of actors $i$ and $j$ are connected 
in a layer if they have worked in a movie of the corresponding genre. Note that not 
every actor has worked in movies of every genre. Thus the number of nodes may 
differ across the genres. However, since we have considered multiplex network with 
each node in a layer having its counter part in the other layers, we consider only 
those nodes which are common in the movie actor network consisting of two layers.
Fig.~\ref{Fig_fclus_Fnl_bolly} demonstrates change in $f_{clus}^1$ of an 
un-delayed layer with respect to the change in the delay value in 
the other delayed ($\tau_2$) for a layer of a multiplex network consisting of 
layers denoting comedy and crime genre for the 2008-2012 data set. 
Whereas, the delayed layer exhibits 
robustness in the enhancement in CS against changes in the delay value of the
 other layer (Fig.~\ref{Fig_fclus_Fnl_bolly}).

In summary, we have investigated the interplay of multiplexing and communication delays on cluster synchronization in multiplex networks,
and elucidated several nontrivial (and anti-intuitive) results. Namely, we have first shown that delayed interactions in a layer enhance or suppresse CS of the other layers, depending on the parity of the delay value. Second, and in contrast to the un-delayed case, we gave evidence that CS of a regular network gets always enhanced upon multiplexing it with a delayed network, which may display arbitrary architecture and delay values.
As a third result, we have shown that enhancement in CS is maximum if the mirror (replica) nodes have the same degree in both layers, while a mismatch in the degree-degree correlation of the mirror nodes leads to the suppression in CS of a un-delayed sparse network.
All these results suggest that two cases have to be distinguished: a un-delayed layer coupled with delayed layers, and a delayed layer multiplexing with other (delayed or un-delayed) layers. In the former case, the enhancement of CS is susceptible to the delay value and the connection density of the other layers, in the latter CS is instead always enhanced.
Therefore, the overall conclusion is that CS of a layer in a multiplex network can be properly tuned by the delay value in the other layers.
Applications of this result are actually relevant, as earlier works have shown that various properties, particularly, the anti-correlation of mirror nodes
make a network vulnerable to the random failure \cite{MD,K.-1_goh_multiplex1}. Our results suggest, instead, that the synchronization in one layer can be made robust against changes in the delay value (as well as network architecture) of the other layer. These results are therefore of value for  multiplex systems with more than one type of vital resources, as for instance water supply and power plants of a city \cite{book_grid,K.-1_goh_multiplex2}, where controlling synchronizability is a task of fundamental importance.

SJ acknowledges DST (EMR/2014/000368) grants for financial support.

\end{document}